\documentclass[english,aps,prd,nofootinbib,superscriptaddress,12pt]{revtex4-1}

\usepackage{graphicx} 

\usepackage[T1]{fontenc}
\usepackage{lmodern}
\usepackage{caption}
\usepackage{floatrow}

\usepackage{subcaption}
\usepackage{hyperref}
\hypersetup{colorlinks,linkcolor={blue},citecolor={blue},urlcolor={blue}}
\usepackage[normalem]{ulem}

\usepackage{color}

\usepackage{mathtools}
\usepackage{physics}
\usepackage{amsmath}
\usepackage{amssymb}
\usepackage{slashed}

\begin{document}

\title{Neutrino Lorentz invariance violation and ultralight axion-like dark matter searches at the European Spallation Source}   

\date{\today}

\author{Rub\'en Cordero}\email{rcorderoe@ipn.mx} \affiliation{Departamento de F\'{\i}sica, Escuela Superior de F\'{\i}sica y Matem\'aticas del Instituto Polit\'ecnico Nacional, Unidad Adolfo L\'opez Mateos, Edificio 9, 07738 Ciudad de M\'exico, Mexico}
\author{Luis A. Delgadillo} \email{ldelgadillof2100@alumno.ipn.mx}
\affiliation{Departamento de F\'{\i}sica, Escuela Superior de F\'{\i}sica y Matem\'aticas del Instituto Polit\'ecnico Nacional, Unidad Adolfo L\'opez Mateos, Edificio 9, 07738 Ciudad de M\'exico, Mexico}
\affiliation{Institute of High Energy Physics, Chinese Academy of Sciences, Beijing 100049, China} 

\begin{abstract}
\noindent
The dark matter conundrum stands out as one of the central challenges to understand in both particle physics and cosmology. The ultralight axion-like particle (UALP) is an appealing dark matter candidate that can be searched for in neutrino oscillation experiments. In this work, we examine the interaction among neutrinos and a UALP; such interaction might induce violations of the Lorentz and $CPT$ symmetries. We assess the sensitivity to both the isotropic $CPT-$odd Standard Model Extension (SME) coefficients $(a_{L})^T$ and effective neutrino-UALP couplings $\tilde{g}$ at the next-to-next generation neutrino oscillation experiment at the European Spallation Source (ESS).
\end{abstract}

\maketitle




\section{Introduction}
\label{intro}

One of the greatest mysteries of nature that has to be resolved is the dark matter dilemma. A plethora of proposals have been suggested in the literature. As an illustration, consider Weakly Interacting Massive Particles (WIMPs), including neutralinos, gravitinos, and sneutrinos in supersymmetric extensions of the Standard Model (SM)~\cite{Arbey:2021gdg}, as well as ultralight, weakly interacting particles such as axions, which, as in the case of WIMPs, are appealing dark matter proposals due to their theoretical motivation and potential observational and experimental prospects to search for them~\cite{Arias:2012az,Ringwald:2013via}.

The axion (a neutral pseudo-scalar boson) was postulated as a solution to the so-called strong $CP$ problem in the theory of strong interactions QCD~\cite{Peccei:1977hh, Weinberg:1977ma, Wilczek:1977pj, Kim:1979if}.
For reviews on axion and ultralight axion-like particles (ALPs) as dark matter and cosmology, see, e.g., Refs.~\cite{Sikivie:2006ni, Ringwald:2013via, Marsh:2015xka, OHare:2024nmr}. 
An ALP is phenomenologically similar to the QCD axion, without solving the strong $CP$ problem~\cite{OHare:2024nmr}. As in the case of the QCD axion, ultralight ALPs can behave as cosmological dark matter~\cite{Hui:2016ltb}, and, in some scenarios, they can generate the correct dark matter (DM) abundance~\cite{Arias:2012az, Marsh:2015xka, OHare:2024nmr}. For instance, several ultralight scalars appear in some string theories~\cite{Witten:1984dg, Choi:1985je, Svrcek:2006yi, Arvanitaki:2009fg, Acharya:2010zx, Cicoli:2012sz, Cicoli:2021gss}. Some searches of ALPs as dark matter include: 21 cm intensity mapping (IM)~\cite{Vanzan:2023gui, Bauer:2020zsj}, gravitational waves~\cite{Michimura:2021hwr, Tsutsui:2022zos}, electromagnetic radiation~\cite{Arza:2019nta}, Lyman-$\alpha$ forest~\cite{Rogers:2020ltq}, velocity acoustic oscillations (VAOs)~\cite{Hotinli:2021vxg}, Hubble and Webb galaxy UV luminosities~\cite{Winch:2024mrt}, axion star explosions~\cite{Escudero:2023vgv}, dwarf galaxies~\cite{Todarello:2023hdk}, time-dependent magnetic fields~\cite{Arias:2016zqu}, pulsar timing array~\cite{Guo:2023hyp}, as well as solid-state nuclear magnetic resonance~\cite{Aybas:2021nvn}.~\footnote{Further searches of ultralight scalars as DM candidates include atomic clocks~\cite{Arvanitaki:2014faa}, resonant-mass detectors~\cite{Arvanitaki:2015iga}, axion-photon couplings~\cite{Arza:2022dng, Yi:2022fmn, Arza:2023rcs, Gao:2023und}, and atomic gravitational wave detectors~\cite{Arvanitaki:2016fyj}.}

As far as searches of ALPs as DM in the neutrino sector are concerned, in Ref.~\cite{Huang:2018cwo}, the authors studied the phenomenology and astrophysical constraints derived from a neutrino interaction with a UALP as dark matter, the authors set limits on the neutrino$-$UALP couplings from the oscillations of neutrinos at several experimental configurations, including DUNE. Furthermore, the implications of this type of coupling with an axion-like particle as DM (with mass $m\sim 10^{-22}$ eV), were studied in the context of high-energy neutrinos from extra-galactic sources~\cite{Reynoso:2022vrn}. Recently, there has been an interest in searches of neutrino non-standard interactions with scalar fields at the ESSnuSB experiment~\cite{Cordero:2022fwb, ESSnuSB:2023lbg}. Besides, in Ref.~\cite{Fierlinger:2024aik}, the authors explore the potential to search for ultralight axion-like dark matter interactions with neutrons produced at the European Spallation Source.

From a phenomenological point of view, it has been argued that potential Lorentz invariance violations (LIV) and violations of the $CPT$ symmetry in the neutrino sector could emerge from neutrino non-standard interactions with scalar fields~\cite{Gu:2005eq, Ando:2009ts, Klop:2017dim, Capozzi:2018bps, Farzan:2018pnk, Ge:2019tdi, Smirnov:2019cae, Gherghetta:2023myo, Lambiase:2023hpq, Cordero:2023hua, Cordero:2024hjr}. For some reviews regarding Lorentz invariance and $CPT$ violations in the neutrino sector, we refer the reader to Refs.~\cite{Kostelecky:2003cr, Diaz:2016xpw, Torri:2020dec, Moura:2022dev}. Besides, for some studies regarding potential violations of the Lorentz and $CPT$ symmetries in neutrino oscillation experiments, see, e.g., Refs.~\cite{Barenboim:2018ctx, IceCube:2021tdn, Rahaman:2021leu, Sahoo:2021dit, Agarwalla:2023wft, Raikwal:2023lzk, Barenboim:2023krl, Pan:2023qln}. In this paper, we study the sensitivity to the isotropic $CPT-$odd SME coefficients $(a_{L})^T$ and effective neutrino-UALP couplings $\tilde{g}$ at the proposed European Spallation Source neutrino Super Beam (ESSnuSB) experiment. Furthermore, we outline some cosmological implications of the neutrino - ultralight axion-like dark matter interaction.

The manuscript is organized as follows: In Section~\ref{framework}, we introduce the theoretical framework and phenomenology of Lorentz invariance violation and the neutrino - ultra-light axion-like dark matter coupling. In Section~\ref{cosmo}, we discuss some cosmological aspects of neutrino$-$UALP interaction. In Section~\ref{method}, we describe the experimental configuration and methodology regarding searches of the aforementioned scenarios in the ESSnuSB proposal. In Section~\ref{results}, we show our results of the potential signatures of Lorentz violation and ultra-light axion-like dark matter couplings via oscillations of neutrinos at the ESSnuSB experiment. Finally, in Section~\ref{conclusion}, we give our conclusions and comment on future assessments in this direction.


\section{Theory Preliminaries}
\label{framework}

Within the standard model extension (SME) framework~\cite{Colladay:1998fq}, LIV effects in the fermion sector are parameterized by the effective Lagrangian~\cite{Barenboim:2022rqu}
\begin{equation}
    \mathcal{L}_{ \text{eff}}^{\text{Fermion}} = \mathcal{L}_{\text{SM}}^{\text{Fermion}} + \mathcal{L}_{\text{LIV}} + \text{h.c.},
\end{equation}
\begin{equation}
    \mathcal{L}_{\text{LIV}}= - \frac{1}{2} \Big\{ p^{\mu}_{\alpha \beta}\Bar{\psi}_{\alpha} \gamma_{\mu} \psi_{\beta} + q^{\mu}_{\alpha \beta}\Bar{\psi}_{\alpha} \gamma_{5} \gamma_{\mu} \psi_{\beta} -i r^{\mu \nu}_{\alpha \beta}\Bar{\psi}_{\alpha} \gamma_{\mu}  \partial_{\nu}  \psi_{\beta}- i s^{\mu \nu}_{\alpha \beta}\Bar{\psi}_{\alpha} \gamma_{5} \gamma_{\mu} \partial_{\nu} \psi_{\beta} \Big\},
\end{equation}
in the case of neutrinos, it is convenient to define
\begin{equation}
    (a_{L})^{\mu}_{\alpha \beta} = (p+q)^{\mu}_{\alpha \beta}, ~~~ (c_{L})^{\mu \nu}_{\alpha \beta} = (r+s)^{\mu \nu }_{\alpha \beta},
\end{equation}
the $(a_{L})^{\mu}_{\alpha \beta}$ coefficients are $CPT$-odd and break the $CPT$ symmetry, while the $ (c_{L})^{\mu \nu}_{\alpha \beta}$ coefficients are $CPT$-even, preserving the $CPT$ symmetry~\cite{Kostelecky:2003cr, Sahoo:2021dit}.

Nevertheless, if we are interested in examining the isotropic $(\mu = \nu = T)$ LIV contributions at neutrino oscillation experiments, either the $CPT$-even SME coefficients $(c_L)^{TT}_{\alpha \beta}=c_{\alpha \beta}$ or the $CPT$-odd SME coefficients $(a_L)^{T}_{\alpha \beta}=a_{\alpha \beta}$, we can consider the corresponding Hamiltonian
\begin{equation}
    H=H_{\text{vacuum}} + H_{\text{matter}} + H_{\text{LIV}},
\end{equation}
here, $H_{\text{vacuum}}$ and $H_{\text{matter}}$ represent the neutrino Hamiltonian in vacuum and matter, respectively. Additionally, the contribution from the LIV sector is
\begin{equation}
\label{LIVHAM}
 H_{\text{LIV}} =   \left(
\begin{array}{ccc}
 a_{ee} & a_{e \mu} & a_{e \tau} \\
 a_{e \mu}^{*} & a_{\mu \mu} &  a_{ \mu \tau} \\
 a_{e \tau}^{*} & a_{ \mu \tau}^{*} & a_{ \tau \tau} \\
\end{array}
\right) - \frac{4}{3} E_\nu \left(
\begin{array}{ccc}
 c_{ee} & c_{e \mu} & c_{e \tau} \\
 c_{e \mu}^{*} & c_{\mu \mu} &  c_{ \mu \tau} \\
 c_{e \tau}^{*} & c_{ \mu \tau}^{*} & c_{ \tau \tau} \\
\end{array}
\right),
\end{equation}
being $E_\nu$ the neutrino energy, $a_{\alpha \beta} = |a_{\alpha \beta}| e^{i \phi_{\alpha \beta}}$, and $c_{\alpha \beta} = |c_{\alpha \beta}| e^{i \phi_{\alpha \beta}^c}$, accordingly. The factor $-4/3$ arises from the non-observability of the Minkowski trace of the $CPT$-even SME coefficients ($c_{L})^{\mu \nu}_{\alpha \beta}$~\cite{Barenboim:2022rqu, Sahoo:2021dit}.

\subsection{Impact of LIV parameters on the oscillation probability}
\label{proba}

In this subsection, we consider the impact of the isotropic LIV parameters $a_{\alpha \beta} = |a_{\alpha \beta}| e^{i \phi_{\alpha \beta}}$ on the neutrino oscillation probability in the context of the ESSnuSB experimental proposal. To illustrate such LIV effects at the probability level, we have employed the analytical expressions for both $P(\nu_\mu \rightarrow \nu_e)$ appearance channel and $P(\nu_\mu \rightarrow \nu_\mu)$ disappearance channel, as given in Section II of Ref.~\cite{Agarwalla:2023wft}.

\subsubsection{Electron neutrino appearance channel}

As discussed in Refs.~\cite{Agarwalla:2023wft, Raikwal:2023lzk}, the electron neutrino appearance probability is mostly sensitive to LIV effects from the $e-\mu$ and $e-\tau$ sectors, accordingly. Modifications of the standard neutrino oscillation probability (SM) in the presence of LIV parameters, will show a significant deviation from the SM case depending on the value of the associated phase $\phi_{\alpha \beta}$.

\begin{figure}[H]
\begin{subfigure}[h]{0.49\textwidth}
			\caption{  }
			\label{fpa}
\includegraphics[width=\textwidth]{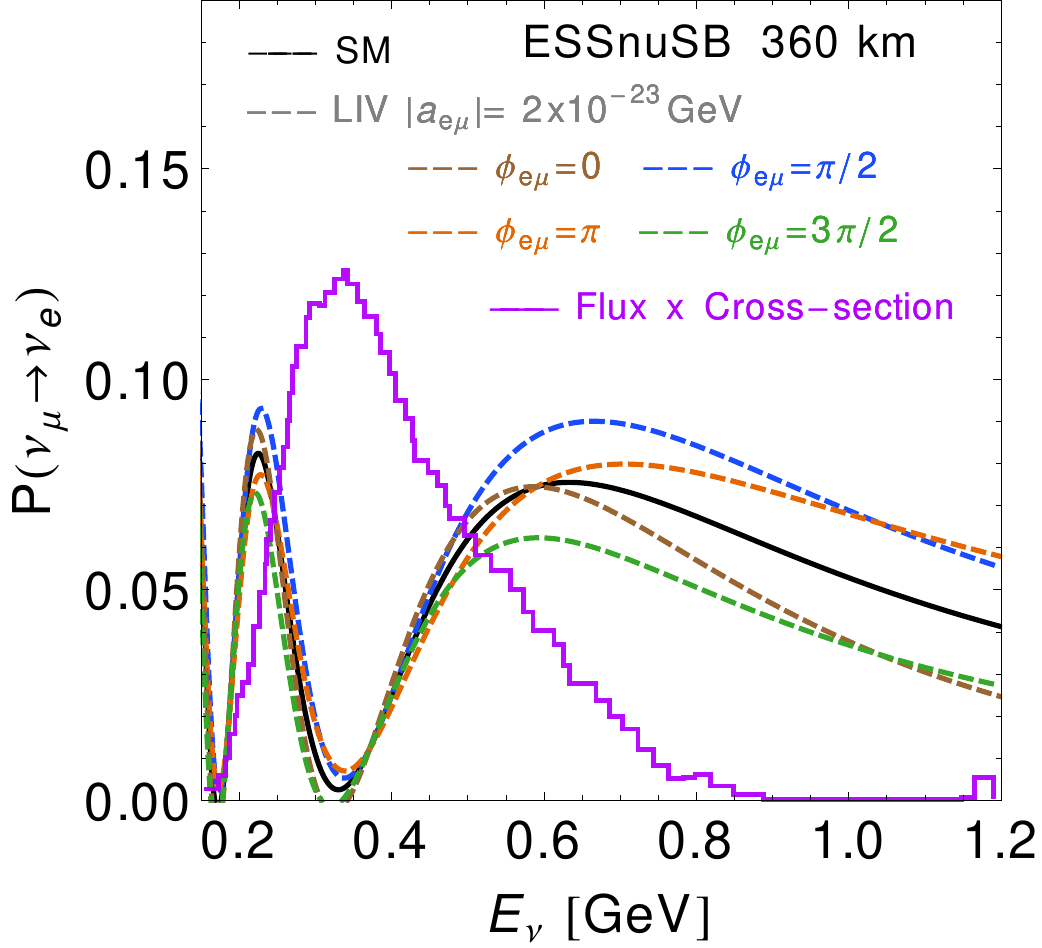}
		\end{subfigure}
		\hfill
		\begin{subfigure}[h]{0.49\textwidth}
			\caption{}
			\label{fpb}
	\includegraphics[width=\textwidth]{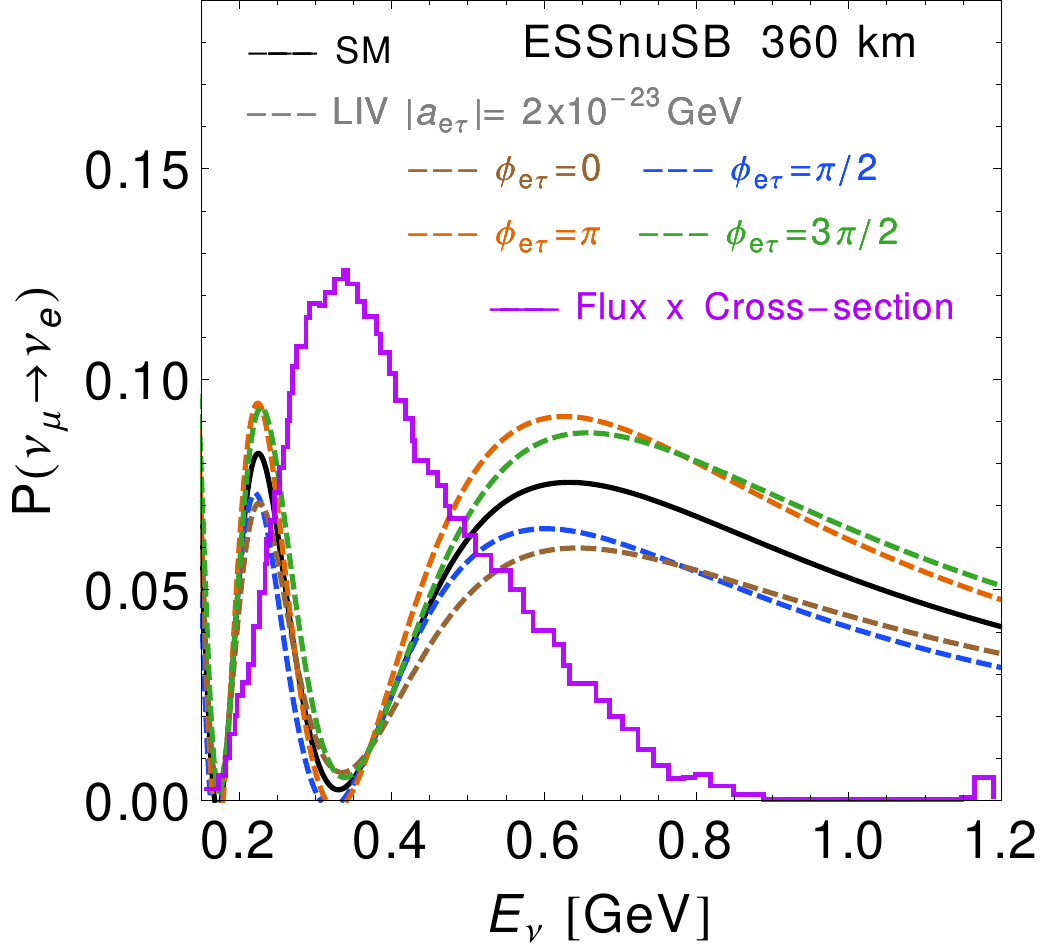}
		\end{subfigure}
		\hfill	
		 \caption{$\nu_\mu \rightarrow \nu_e$ appearance probability as a function of the neutrino energy $E_\nu$ in the presence of LIV from the $e-\mu$ (left panel) and $e-\tau$ (right panel) sectors, respectively. All the oscillation parameters are fixed to their NO best-fit values~\cite{deSalas:2020pgw}.}
  \label{fpess}
\end{figure}

The left panel of Fig.~\ref{fpess} considers the impact of LIV on the $P(\nu_\mu \rightarrow \nu_e)$ oscillation probability from the $e-\mu$ sector. We have fixed the magnitude $|a_{e \mu}| = 2 \times 10^{-23}$ GeV and chosen four values of the LIV phases $\phi_{e \mu}$: $(\phi_{\mu \tau} = 0,~\pi/2,~\pi,~3 \pi/2)$ displayed in (brown, blue, orange, green)-dashed lines, accordingly. Furthermore, the purple contour shows the neutrino flux times cross-section at the electron neutrino appearance channel as presented in Fig.~1 of Ref.~\cite{ESSnuSB:2021azq}.

\subsubsection{Muon neutrino disappearance channel}

On the other hand, the muon neutrino disappearance probability is mostly sensitive to LIV effects from the $\mu-\tau$ sector~\cite{Agarwalla:2023wft}.
\begin{figure}[H]
\begin{subfigure}[h]{0.49\textwidth}
			\caption{  }
			\label{fa}
\includegraphics[width=\textwidth]{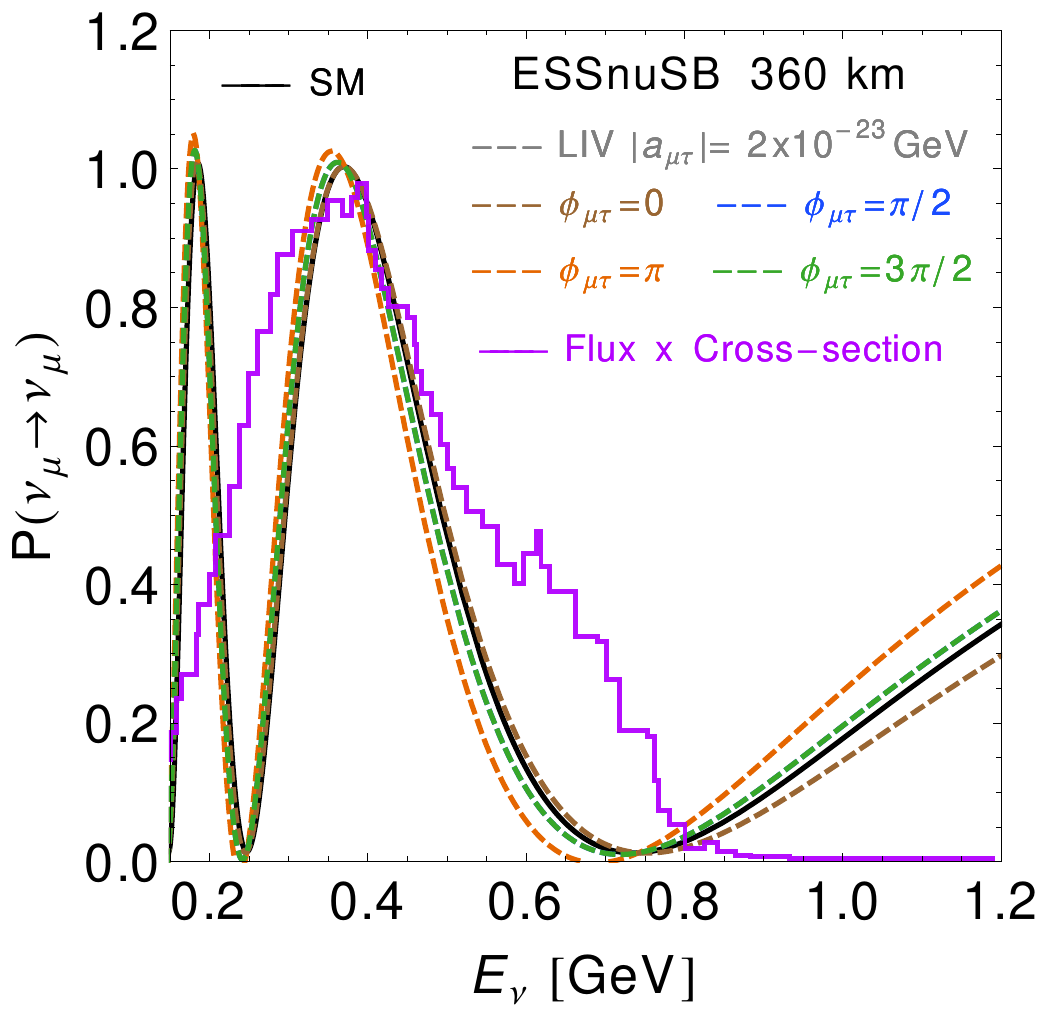}
		\end{subfigure}
		\hfill
		\begin{subfigure}[h]{0.49\textwidth}
			\caption{}
			\label{fb}
	\includegraphics[width=\textwidth]{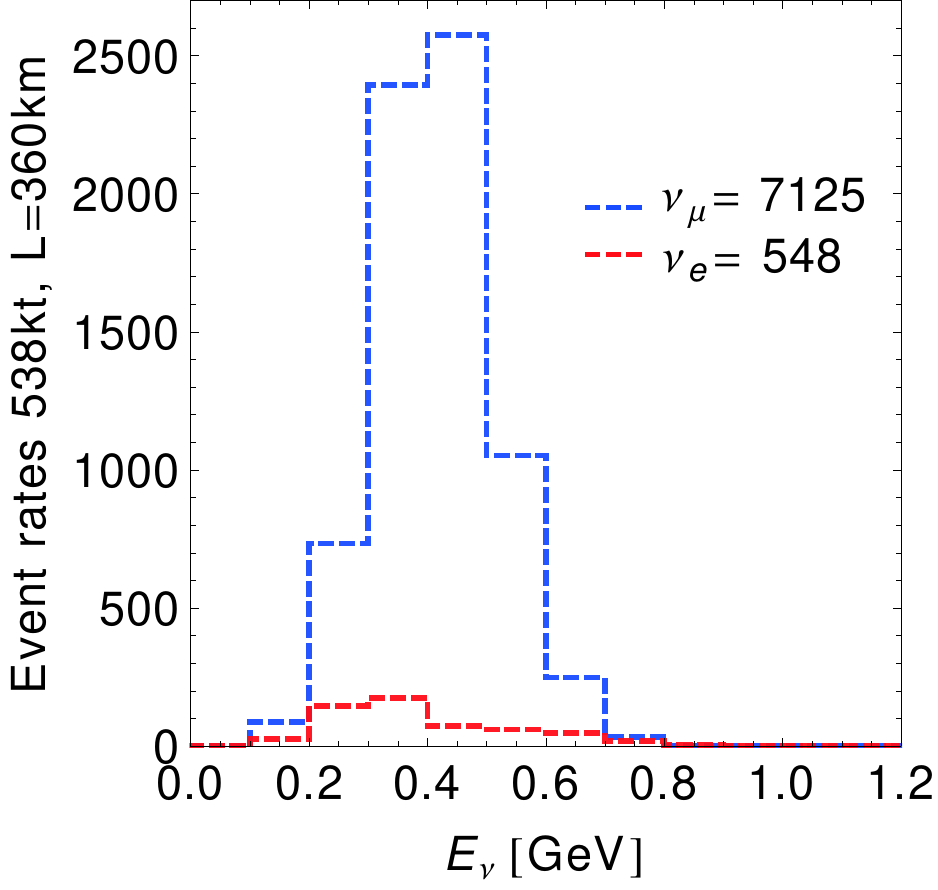}
		\end{subfigure}
		\hfill	
		 \caption{$\nu_\mu \rightarrow \nu_\mu$ disappearance probability as a function of the neutrino energy $E_\nu$ in the presence of LIV from the $\mu-\tau$ sector (left panel). Expected reconstructed electron neutrino ($\nu_e$) and muon neutrino ($\nu_\mu$) events at the ESSnuSB setup (right panel). All the oscillation parameters are fixed to their NO best-fit values~\cite{deSalas:2020pgw} (see Section~\ref{method} for details regarding the ESSnuSB experimental setup and simulation).}
  \label{fess}
\end{figure}

The left panel of Fig.~\ref{fess} considers the impact of LIV on the $P(\nu_\mu \rightarrow \nu_\mu)$ oscillation probability from the $\mu- \tau$ sector. We set the magnitude $|a_{\mu \tau}| = 2 \times 10^{-23}$ GeV and choose four values of the LIV phases $\phi_{\mu\tau}$: $(\phi_{\mu \tau} = 0,~\pi/2,~\pi,~3 \pi/2)$ shown in (brown, blue, orange, green)-dashed lines, accordingly. Moreover, the purple contour displays the neutrino flux times cross-section at the muon neutrino disappearance channel as presented in Fig.~1 of Ref.~\cite{ESSnuSB:2024yji}. Besides, in the right panel of Fig.~\ref{fess}, we show our simulated electron and muon neutrino events, considering the standard three neutrino oscillations picture at the ESSnuSB $L=$ 360 km configuration.

Although the effect of LIV from the $\mu- \tau$ sector appears to be marginal compared with those from the $e-\mu$ and $e-\tau$ sectors (Fig.~\ref{fpess}), the sensitivity to the $a_{\mu \tau}$ parameter will be compensated by larger statistics from the muon neutrino disappearance channel, as displayed in the right panel of Fig.~\ref{fess}.


\subsection{Neutrinophilic axion-like dark matter}
\label{neutrinophilic}

The derivative interaction among neutrinos with a pseudo-scalar axion-like particle (ALP) is given as~\cite{Huang:2018cwo, Reynoso:2022vrn, Gherghetta:2023myo}
\begin{equation}
\label{alplag}
   - \mathcal{L}_{\text{eff}}=  \tilde{g}_{\alpha \beta} \partial_\mu \varphi \bar{\nu}_{L\alpha} \gamma^{\mu}\gamma_5 \nu_{L\beta}~,
\end{equation}
here, $\varphi$ is a pseudo-scalar ALP field, $\tilde{g}_{\alpha \beta} = \tilde{g}_{\beta \alpha}^{*}$ are the effective couplings among the ALP and the left-handed active neutrinos $\nu_L$, and $\alpha, \beta = (e,~\mu,~\tau)$ are flavor indices. For instance, in some models, this type of interaction emerges~\cite{Baek:2019wdn, Baek:2020ovw, Cox:2021lii}.~\footnote{Besides, within the $M$-theory axiverse~\cite{Acharya:2010zx, Marsh:2015xka} an ultralight ($m_\varphi \sim 10^{-15}$ eV) axion-like particle arises, with decay constant $F_{\varphi}\sim 10^{16}$ GeV.} In this scenario, by considering ultra-relativistic neutrinos and $\varphi \simeq\varphi(t)$, an axion-like particle, we can obtain the following relationship among the $CPT-$odd SME coefficients $(a_L)^{\mu}_{\alpha \beta} = a^{\mu}_{\alpha \beta}$, and the effective neutrino$-$ALP interaction~\cite{Lambiase:2023hpq, Huang:2018cwo}
\begin{equation}
\label{livdark}
    a^{\mu}_{\alpha \beta} \bar{\nu}_{L\alpha} \gamma_{\mu}\gamma^5 \nu_{L\beta} \rightarrow \tilde{g}_{\alpha \beta} 
 \partial^{\mu} \varphi \bar{\nu}_{L\alpha} \gamma_{\mu}\gamma^5 \nu_{L\beta}~,
\end{equation}
the pseudo-scalar field $\varphi\simeq \varphi(t)$, can be identified as a dark matter candidate (see Sec.~\ref{cosmo}). 
Furthermore, in this scenario, violations of the Lorentz and $CPT$ symmetries emerge via the $CPT-$odd SME coefficients $a_{\alpha \beta}^{\mu} \rightarrow \tilde{g}_{\alpha \beta} \partial^{\mu} \varphi$~\cite{Lambiase:2023hpq}. For the case of $\varphi$ with potential $V(\varphi) \simeq m_\varphi^2 \varphi^2$, locally, this field can be expressed as~\cite{Magana:2012ph, Suarez:2013iw, Berlin:2016woy, Ferreira:2020fam}
\begin{equation}
\label{ualp}
   \varphi(t,X) \simeq \frac{\sqrt{2 \rho_{\varphi, \odot}}}{m_\varphi} \sin \big(m_\varphi (t -\langle v_\varphi \rangle X) \big)~,
\end{equation}
being $\rho_{\varphi, \odot}$ the local ultralight axion-like particle (UALP) field density, $X=(x,y,z)$ the spatial components, and $\langle v_\varphi \rangle$ the virialized UALP velocity (which is of the order $\mathcal{O}(10^{-3})c$, see, e.g., Refs.~\cite{Klop:2017dim, Huang:2018cwo, Cordero:2023hua}) in the Milky Way (MW).

In this study, we will consider a UALP density of $\rho_{\varphi} \simeq  \rho_{\text{DM}} \simeq 2 \times 10^{-11}$ eV$^4$ (locally $\rho_{\text{DM}, \odot} \sim 10^{5} \rho_{\text{DM}}$~\cite{Planck:2018vyg}).~\footnote{From recent analyses, the estimation of the local DM density coincides within a range of $\rho_{\text{DM}, \odot} \simeq 0.3-0.6~ \text{GeV/cm}^3$~\cite{deSalas:2019pee, deSalas:2020hbh, Sivertsson:2022riu}.} Thus, depending on the ALP mass $m_\varphi$, density, and effective couplings $\tilde{g}_{\alpha \beta}$ ($\tilde{g}_{ee} \lesssim 10^{-9}$ GeV$^{-1}$ for electron-ALP couplings~\cite{Reynoso:2022vrn}), for ultra-relativistic neutrinos, one can obtain bounds for the $CPT$-odd SME coefficients $a_{\alpha \beta}^\mu (t)$ as in the case where $\varphi$ is considered to be scalar field dark matter~\cite{Gherghetta:2023myo, Lambiase:2023hpq}.

The effective Lagrangian in Eq.~(\ref{alplag}) could induce violations of Lorentz invariance and $CPT$ at neutrino oscillation experiments via the $CPT$-odd SME coefficients $a^{\mu}_{\alpha \beta} \rightarrow \tilde{g}_{\alpha \beta} \partial^{\mu} \varphi$. For instance, for the case of an ALP field $\varphi$ with mass $m_\varphi \sim 10^{-15}$ eV, hence, at present, the local value of the field amplitude in the MW would be $|\varphi_{\odot}| \sim 10^{12}$ eV, therefore $|\dot{\varphi}_{\odot}| \simeq m_\varphi |\varphi_{\odot}| \simeq \sqrt{2\rho_{\varphi,\odot}} \sim 10^{-21}$ GeV$^2$. Thus, from Eqs.~(\ref{livdark}) and (\ref{ualp}), the isotropic SME coefficients $a_{\alpha \beta}(t)$ are
\begin{equation}
    a_{\alpha \beta} (t) = a_{\alpha \beta} \cos(m_\varphi t) \simeq \tilde{g}_{\alpha \beta} |\dot{\varphi}_{ \odot}| \cos(m_\varphi t).
\end{equation}
Comparably to the situation examined in~\cite{Klop:2017dim}, directional-dependent contributions ($a_{\alpha \beta}^{X}$) will be sub-leading due to the small virialized UALP velocity $\langle v_\varphi \rangle$
\begin{equation}
    |a_{\alpha \beta}^{X}|\simeq \tilde{g}_{\alpha \beta}|\nabla \varphi_{\odot} \cdot \hat{p}_\nu| \simeq \tilde{g}_{\alpha \beta} m_\varphi |\phi_{\odot} \langle v_\varphi \rangle \hat{v}_\varphi \cdot \hat{p}_\nu| \sim \tilde{g}_{\alpha \beta} \sqrt{2\rho_{\varphi,\odot}} \langle {v}_\varphi\rangle \sim 10^{-3}|a_{\alpha \beta}|~,
\end{equation}
these coefficients have been investigated within the framework of active-sterile neutrino oscillations and supernovae neutrino emission~\cite{Lambiase:2023hpq}. 


\section{Cosmological aspects of the neutrino$-$UALP interaction}
\label{cosmo}

At the cosmological level, the evolution of the interacting ALP field $\varphi$ with neutrinos can be described through energy exchange among the two fluids~\cite{Boehmer:2008av, Chongchitnan:2008ry}
\begin{equation}
    \dot{\rho}_{\nu} +4H\rho_\nu =Q,
\end{equation}
\begin{equation}
    \dot{\rho}_{\varphi} +3H(\rho_\varphi + P_\varphi) =-Q,
\end{equation}
where the explicit form of the energy exchange term, $Q$, is model dependent.~\footnote{For instance, in the scenario explored by the authors of Ref.~\cite{Simpson:2016gph}, the energy exchange term is $Q= \tilde{g} n_\nu \ddot{\phi}$, being $\tilde{g}$ an effective coupling constant, $\phi$ a scalar field, and $n_\nu$ the neutrino number density. Within this scenario, $\phi$ can behave as dark energy~\cite{Simpson:2016gph}. However, if the energy exchange term is $Q \propto \dot{\phi}$, the scalar field could act as dark matter~\cite{Cordero:2023hua}.} 
In addition, the evolution of the Hubble parameter $H$, can be obtained from the Friedmann equation\textbf{}
\begin{equation}
    H^2 = \frac{8\pi G}{3}(\rho_r+ \rho_b+ \rho_\varphi + \rho_\Lambda),
\end{equation}
here, $\rho_r$, $\rho_b$, and $\rho_\Lambda$, are the energy densities associated with radiation, nonrelativistic baryonic matter, and dark energy, respectively.

Furthermore, the evolution of the ALP field $\varphi$ can be determined from its equation of motion
\begin{equation}
\ddot{\varphi}+3H\dot{\varphi}+\frac{dV(\varphi)}{d\varphi}= -\frac{Q}{\dot{\varphi}}~,
    \label{fievolution}
\end{equation}
being $H = \dot{a}/a$ the Hubble parameter, and $a$ the scale factor of the Universe. However, the corresponding energy exchange term $Q$, for the case of the neutrino$-$UALP interaction would be~\cite{Wetterich:1994bg}
\begin{equation}
    Q \simeq \frac{d}{dt} \left(\partial_0 \varphi \langle \bar{\nu}_{L\alpha} \gamma^0  \gamma^5 \nu_{L\beta} \rangle \right) \simeq \frac{d}{dt} \left( \dot{ \varphi} \biggr \langle \frac{ \Vec{\sigma }_\nu \cdot \Vec{p}_\nu }{m_\nu} \biggr \rangle \right) \simeq 0~,
\end{equation}
where $\Vec{\sigma }_\nu$ is the average spin of the neutrinos~\cite{Athar_Singh_2020}, and $\Vec{\sigma }_\nu \simeq 0$~\cite{He:1998ng}, unless we consider a neutrino beam with a particular spin orientation. Furthermore, the equation of motion for the ALP field would be
\begin{equation}
\ddot{\varphi}+3H\dot{\varphi}+\frac{dV(\varphi)}{d\varphi}= - \frac{Q}{\dot{\varphi}} \simeq 0.
\end{equation}
Therefore, in this scenario, the ALP may evolve as a free particle and could be considered a DM candidate, e.g., ultralight axion-like dark matter UALP, with mass $m_\varphi$ in the range $10^{-24}~\text{eV} \lesssim m_\varphi \lesssim 1$ eV~\cite{Arias:2012az, Ferreira:2020fam, Reynoso:2022vrn}.


\section{Experimental setup and methodology}
\label{method}

Long-baseline neutrino oscillation experiments play a central role in both elucidating the mysteries within the standard three-neutrino oscillation paradigm and exploring other novel physics scenarios, including the potential breaking of the Lorentz and $CPT$ symmetries as well as the possibility of ultralight dark matter interactions with neutrinos. 

In this section of the manuscript, we will focus on a long-baseline experimental configuration, the European Spallation Source neutrino Super Beam (ESSnuSB) initiative~\cite{ESSnuSB:2013dql}, which is a proposed next-to-next generation accelerator-based neutrino oscillation experiment at the European Spallation Source (ESS), envisioned to determine and measure the leptonic $CP-$violating phase $\delta_{CP}$~\cite{ESSnuSB:2013dql, Blennow:2019bvl}. It will consist of a water Cerenkov detector with a total mass of up to 538 kt, placed inside a deep mine in Sweden~\cite{ESSnuSB:2013dql, ESSnuSB:2021azq, Alekou:2022emd, Blennow:2019bvl, Ghosh:2023sfi}. Moreover, within the ESSnuSB$+$ proposal~\cite{Giarnetti:2023pkz}, a monitored neutrino beam (based on the ENUBET technology~\cite{Longhin:2014yta}) to improve neutrino cross-section measurements at the ESS is contemplated~\cite{Terranova:2023gie}. 

From the ESSnuSB initiative, two potential locations have been investigated: $L \simeq$ 540 km (Garpenberg mine) and $L \simeq$ 360 km (Zinkgruvan mine)~\cite{Blennow:2019bvl}, however, in this work, we will consider the configuration at a distance of $L \simeq$ 360 km from the beam source (on-axis) at Lund~\cite{ESSnuSB:2021azq, Alekou:2022emd, ESSnuSB:2023ogw, ESSnuSB:2023lbg}. Moreover, within this experimental setup, the mean neutrino energy would be around $E_\nu \sim$ 0.4 GeV~\cite{ESSnuSB:2013dql, Blennow:2019bvl, ESSnuSB:2021azq, Alekou:2022emd, ESSnuSB:2023ogw}. Searches of beyond the Standard Model (BSM) scenarios at ESSnuSB include sterile neutrinos~\cite{KumarAgarwalla:2019blx, Ghosh:2019zvl}, source and detector non-standard neutrino interactions (NSI)~\cite{Blennow:2015nxa}, neutrino decay~\cite{Choubey:2020dhw}, non-unitary lepton mixing~\cite{Capozzi:2023ltl, Chatterjee:2021xyu}, leptonic flavor models~\cite{Blennow:2020snb, Ahn:2022ufs}, ultralight scalar dark matter~\cite{Cordero:2022fwb}, scalar mediated neutrino non-standard interactions~\cite{ESSnuSB:2023lbg}, as well as quantum decoherence in neutrino oscillations~\cite{ESSnuSB:2024yji}.

In order to obtain sensitivities to the LIV coefficients at an ESSnuSB-like experimental setup, we use the \textsc{GLoBES} software~\cite{Huber:2004ka, Huber:2007ji} and its additional NSI tool \emph{snu.c}~\cite{Kopp:2006wp, Kopp:2007rz} which was modified to implement the $CPT-$odd coefficients of the SME at the Hamiltonian level. Moreover, we use the ESSnuSB experimental configuration as developed in Refs.~\cite{Cordero:2022fwb, Delgadillo:2023lyp}, which considers a $10-$year running time, evenly distributed among neutrino and antineutrino modes.  
In this analysis, we have considered the following Hamiltonian, $\tilde{H}(t)$, which in the presence of the $CPT$-odd LIV coefficients is
\begin{equation}
    \tilde{H}(t) = H_0 +V_{\text{MSW}} + a_{\alpha \beta}(t)~,
\end{equation}
where $H_0$ is the neutrino Hamiltonian in vacuum, $V_{\text{MSW}}$ is the MSW potential, and the last term corresponds to the $CPT$-odd LIV Hamiltonian 
\begin{equation}
    a_{\alpha \beta}(t) = \tilde{g}_{\alpha \beta}  \frac{\partial \varphi(t, X)}{\partial t}\simeq \tilde{g}_{\alpha \beta} \sqrt{2\rho_{\varphi,\odot}}\cos(m_\varphi t_\varphi).
\end{equation}
Besides, for simplicity, we will consider an approach performed by the authors of Ref.~\cite{Arguelles:2024cjj}, such that we will fix $\cos(m_\varphi t_\varphi) =1$. Therefore, the isotropic $CPT-$odd SME coefficients will be related to the effective neutrino$-$UALP couplings as $a_{\alpha \beta} \simeq \tilde{g}_{\alpha \beta}\sqrt{2\rho_{\varphi,\odot}}\,$, to obtain sensitivities for the effective neutrino$-$UALP couplings, $\tilde{g}_{\alpha \beta}$, we consider a local DM density of $\rho_{\varphi,\odot} = 2\times 10^{-6}$ eV$^4$. For instance, at the ESSnuSB configuration ($L = 360$ km), the expected ultralight field mass sensitivity is $2.0 \times 10^{-23}~\text{eV} \leq m_{\varphi}^{360~\text{km}} \leq 8.5 \times 10^{-15}~\text{eV}$~\cite{Cordero:2022fwb}. However, we should point out that the sensitivities derived for the effective couplings $\tilde{g}_{\alpha \beta}$, might change when averaging over the ALP oscillation time $t_\varphi$~\cite{Huang:2018cwo}.

Moreover, to quantify the statistical significance of $CPT-$odd SME coefficients $a_{\alpha \beta}$ and effective couplings $\tilde{g}_{\alpha \beta}$, we setup a chi-square test analysis employing both neutrino and anti-neutrino data sets. The total $\chi^2-$function is given as in Ref.~\cite{Cordero:2022fwb}
\begin{equation}
    \chi^2 = \sum_{c} \tilde{\chi}^2_{c} + \chi^2_{\text{prior}}~,
\end{equation}
where the corresponding $\tilde{\chi}^2_{c}-$function stands for each channel $c$, with $c= \big( \nu_{\mu}(\Bar{\nu}_{\mu})\rightarrow \nu_{e} (\Bar{\nu}_{e}),~\nu_{\mu}(\Bar{\nu}_{\mu})\rightarrow \nu_{\mu} (\Bar{\nu}_{\mu}) \big)$, and is provided as in Ref.~\cite{Huber:2002mx}
\begin{equation}
\begin{split}
    &\tilde{\chi}_{c}^2= \min_{\eta_{j}} \Bigg[  \sum _{i}^{n_{\text{bin}}} 2 \Bigg\{ N_{i,\text{test}}^{3 \nu+\text{LIV}}( \Omega, \Theta, \{\eta_{j}\})-N_{i,\text{true}}^{3\nu} + N_{i,\text{true}}^{3\nu} \log \frac{N_{i,\text{true}}^{3\nu}}{N_{i,\text{test}}^{3 \nu+\text{LIV}}( \Omega, \Theta, \{\eta_{j}\})} \Bigg\}\\
    &~~~~~~~~~~~~~~~~+ \sum_{j}^{n_{\text{syst}}} \Big(\frac{\eta_{j}}{\sigma_{j}}\Big)^2 \Bigg]~,
\end{split}
\end{equation}
here, $N_{i, \text{true}}^{3\nu}$ are the simulated events at the $i$-th energy bin, considering the standard three neutrino oscillations framework, $N_{i, \text{test}}^{3\nu +\text{LIV}}( \Omega, \Theta, \{\eta_{j}\})$ are the computed events at the $i$-th energy bin, including $CPT-$odd SME coefficients (one parameter at a time). In addition, $\Omega = \{\theta_{12}, \theta_{13}, \theta_{23}, \delta_{CP}, \Delta m_{21}^2, \Delta m^2_{31}\}$ is the set of neutrino oscillation parameters, while $\Theta = \{|a_{\alpha \beta}|, \phi_{\alpha\beta}, a_{\alpha \alpha}\}$ is the set of isotropic SME coefficients (non-diagonal $a_{\alpha \beta}= |a_{\alpha \beta}| e^{i \phi_{\alpha \beta}}$, diagonal $a_{\alpha \alpha}$), where $\{\eta_{j}\}$ are the nuisance parameters to account for the systematic uncertainties. Furthermore, $\sigma_{j}$ are the systematic uncertainties as described in Section~III of Refs.~\cite{Cordero:2022fwb, Delgadillo:2023lyp}. Besides, to obtain our simulated events, we consider the neutrino oscillation parameters from Ref.~\cite{deSalas:2020pgw} as~\emph{true} values. We show those values in Table~\ref{tab:1}.
In addition, the implementation of external input for the standard oscillation parameters on the $\chi^2$ function is performed via Gaussian priors~\cite{Huber:2002mx}
\begin{equation}
    \chi^2_{\text{prior}}= \sum_{p}^{n_{\text{priors}}}   \frac{\big(\Omega_{p,\text{true}}-\Omega_{p,\text{test}}\big)^2}{\sigma^2_{p}}\,. 
\end{equation}
The central values of the oscillation parameter priors, $\Omega_{p,\text{true}}$, are fixed to their best-fit value from Ref.~\cite{deSalas:2020pgw}, considering the normal mass ordering of neutrino masses, and $\sigma_p$ corresponds to the 1$\sigma$ confidence level (C.L.) of each parameter $p$.
\begin{table}[H]
\caption{\label{tab:1}Standard oscillation parameters used in our analysis~\cite{deSalas:2020pgw}. We consider the normal mass ordering (NO) throughout this study.}
\centering
\begin{tabular}{c  c}
\hline \hline
Oscillation parameter & best-fit \textbf{NO}  \\
\hline 
$\theta_{12}$ & 34.3$^{\circ}$ \\
$\theta_{23}$ & 49.26$^{\circ}$ \\
$\theta_{13}$ &  8.53$^{\circ}$ \\
$\Delta m^2_{21}$ [10$^{-5}$~eV$^2$] & 7.5  \\ 
$|\Delta m_{31}^2|$ [10$^{-3}$~eV$^2$] & 2.55  \\ 
$\delta_{CP}/ \pi$ & 1.08  \\
\hline \hline
\end{tabular}
\end{table}


\section{Results}
\label{results}

In this section, we present our results of the projected sensitivities to the isotropic $CPT-$odd SME coefficients $a_{\alpha \beta}$ and the effective couplings of the neutrino-ultralight axion-like dark matter interaction $\tilde{g}_{\alpha \beta}$,~\footnote{Nevertheless, for relativistic neutrinos, the analysis with ultralight axion-like dark matter, UALP, will be equivalent to the ultralight scalar dark matter (ULDM) case~\cite{Gherghetta:2023myo, Lambiase:2023hpq}.} considering the ESSnuSB 360 km ($L=360$ km) configuration.

\begin{figure}[H]
\begin{subfigure}[h]{0.48\textwidth}
			\caption{  }
			\label{f1a}
\includegraphics[width=\textwidth]{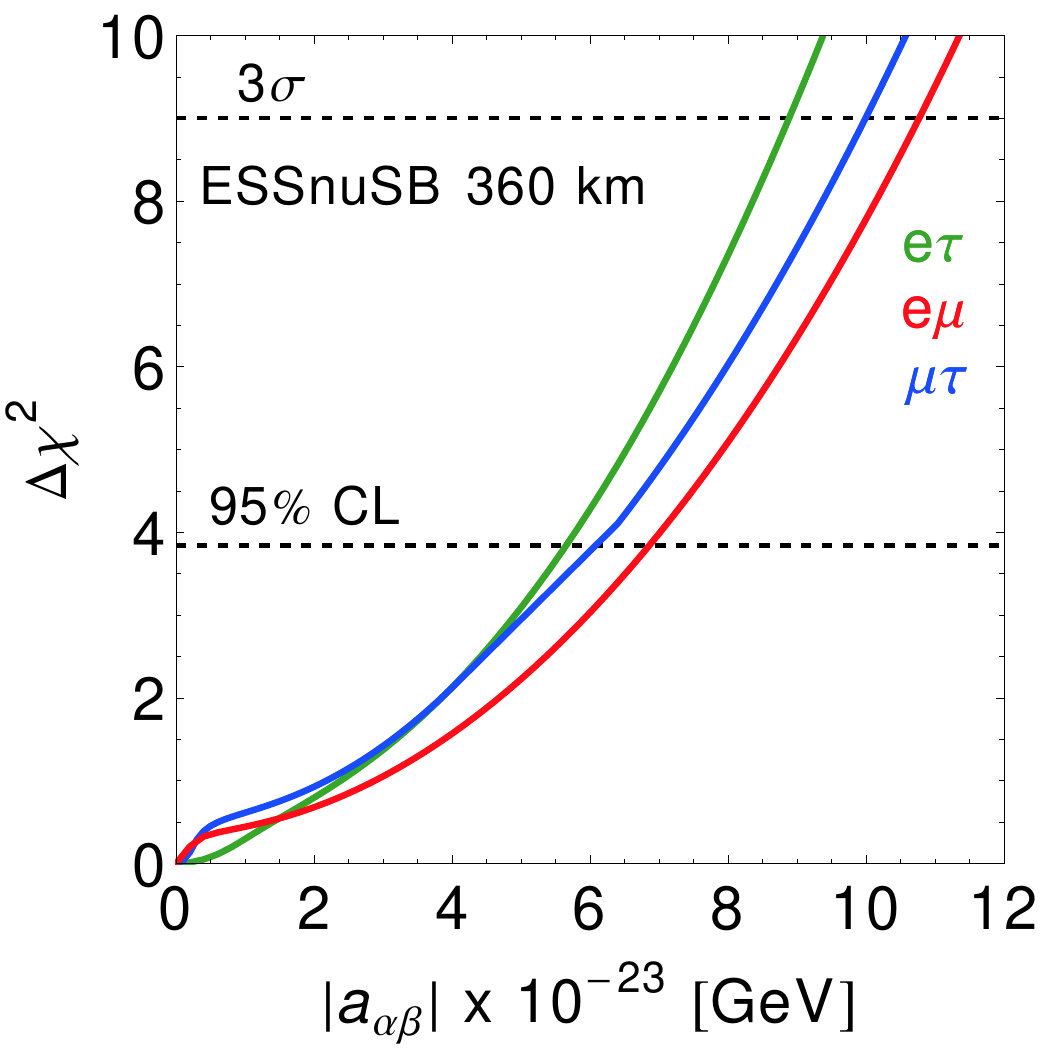}
		\end{subfigure}
		\hfill
		\begin{subfigure}[h]{0.48 \textwidth}
			\caption{}
			\label{f1b}
	\includegraphics[width=\textwidth]{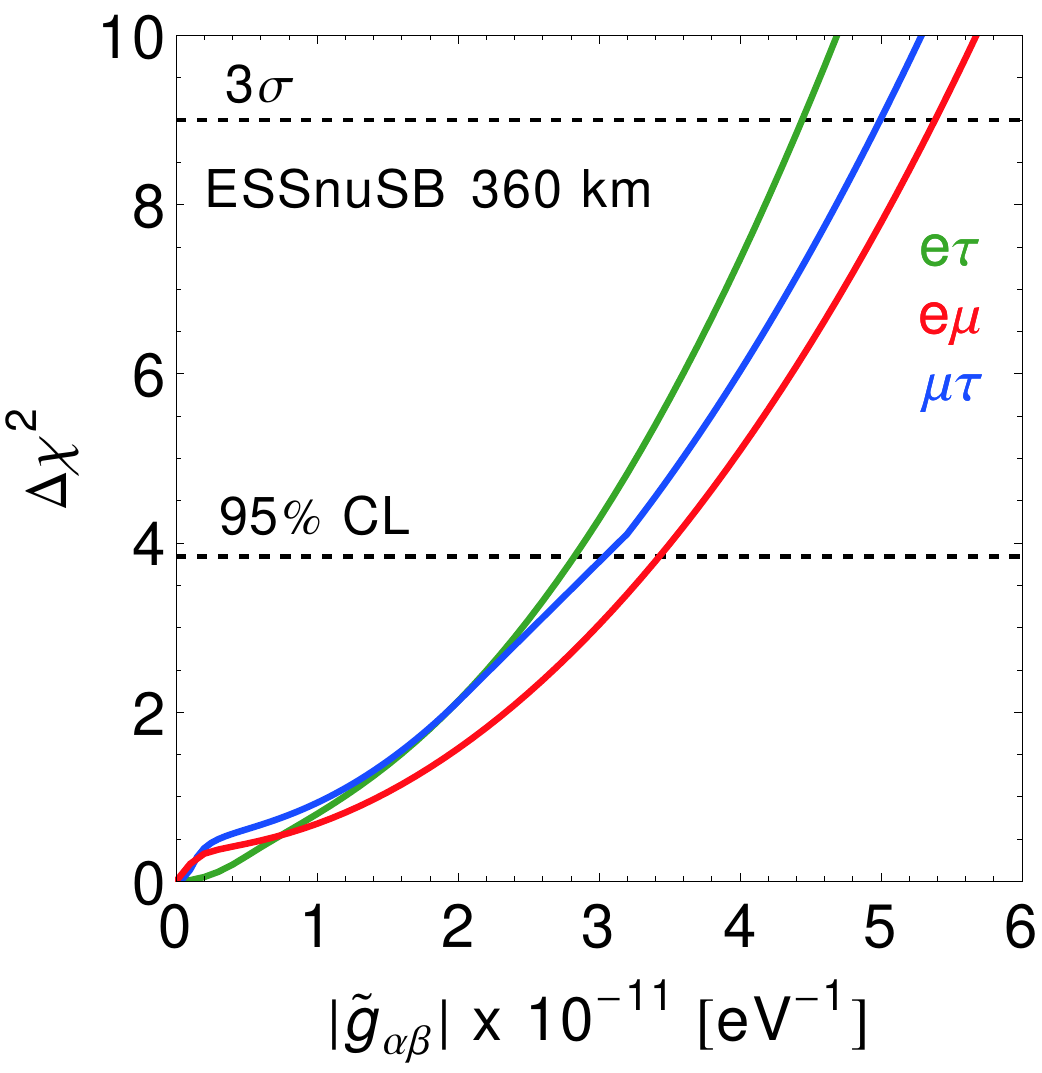}
		\end{subfigure}
		\hfill	
		 \caption{The expected sensitivity regions: SME coefficients $|a_{\alpha \beta}|$ (left panel), and effective couplings $|\tilde{g}_{\alpha \beta}|$ (right panel), accordingly. We have marginalized over the corresponding LIV phases, $\phi_{\alpha \beta}$, in the interval [0$-2\pi$], as well as $\theta_{23}$ and $\delta_{CP}$, around their 1$\sigma$ uncertainty~\cite{deSalas:2020pgw}. All the remaining oscillation parameters are fixed to their NO best fit values. See text for a detailed explanation.}
  \label{f1ess}
\end{figure}
In Fig.~\ref{f1ess}, we display the projected 95\% C.L. and $3\sigma$ sensitivities to the isotropic $CPT-$odd SME coefficients $a_{\alpha \beta}$, and the effective couplings of the neutrino-ultralight axion-like dark matter interaction $\tilde{g}_{\alpha \beta}$, at the ESSnuSB 360 km configuration. The left panel shows the sensitivities to the non-diagonal isotropic SME coefficients: $|a_{\alpha \beta}| \simeq [5-7] \times10^{-23}$ GeV (red, blue, green)-solid lines ($\alpha \beta = e\mu,~\mu \tau,~e \tau$) at 95\% C.L. Here, we have marginalized over the corresponding LIV phases, $\phi_{\alpha \beta}$, in the interval [0$-2\pi$], as well as the standard oscillation parameters $\theta_{23}$ and $\delta_{CP}$. Besides, the right panel displays the sensitivities to the non-diagonal effective neutrino$-$UALP couplings: $|\tilde{g}_{\alpha \beta}|\simeq 3 \times 10^{-11}$ eV$^{-1}$ at 95\% C.L. We observe that the expected sensitivities to these effective couplings at the ESSnuSB setup are comparable with limits ($\tilde{g}_{\alpha \beta} < 2 \times 10^{-11}$ eV$^{-1}$~\cite{Huang:2018cwo}) from the proposed JUNO experiment~\cite{JUNO:2015sjr}.

\begin{figure}[H]
\begin{subfigure}[h]{0.48\textwidth}
			\caption{  }
			\label{f2a}
\includegraphics[width=\textwidth]{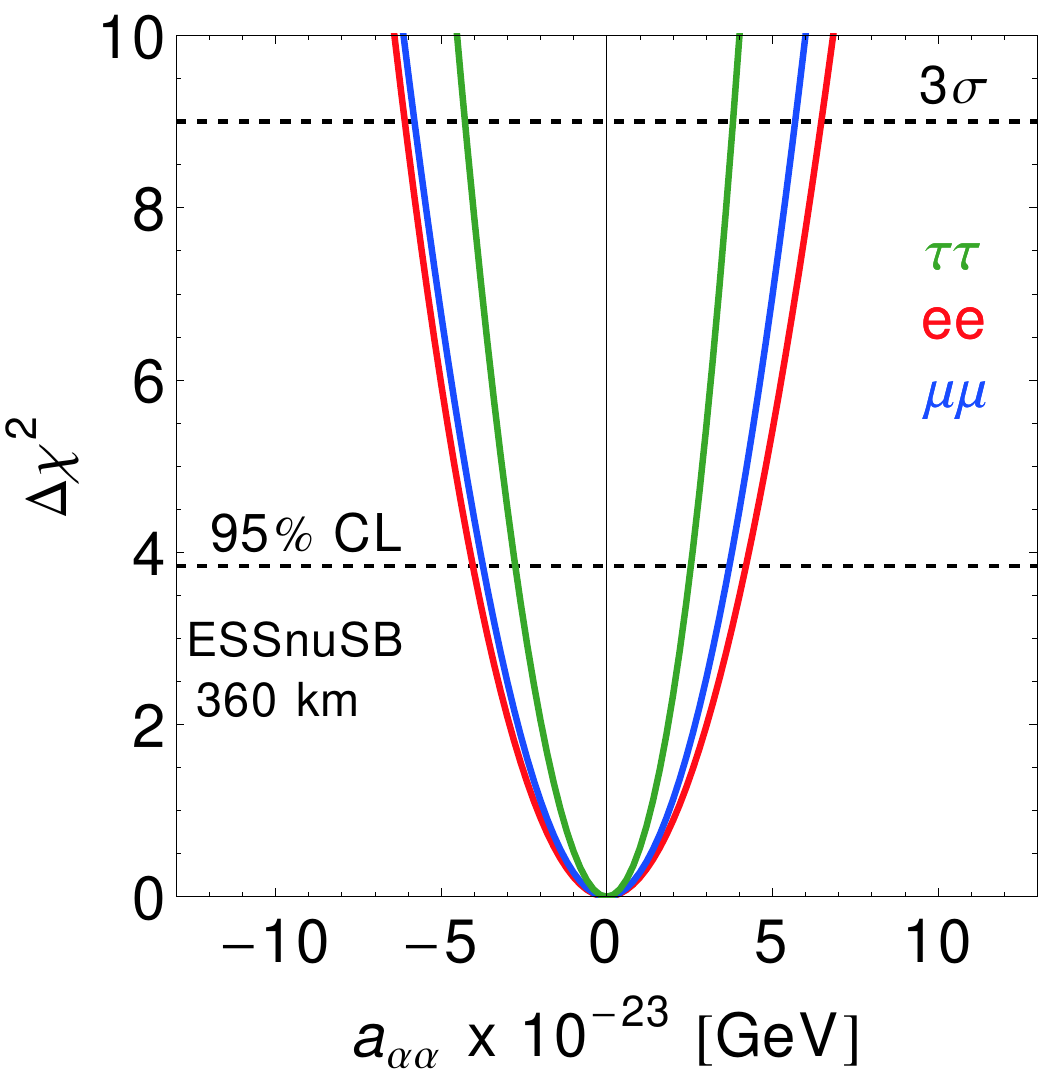}
		\end{subfigure}
		\hfill
		\begin{subfigure}[h]{0.48 \textwidth}
			\caption{}
			\label{f2b}
			\includegraphics[width=\textwidth]{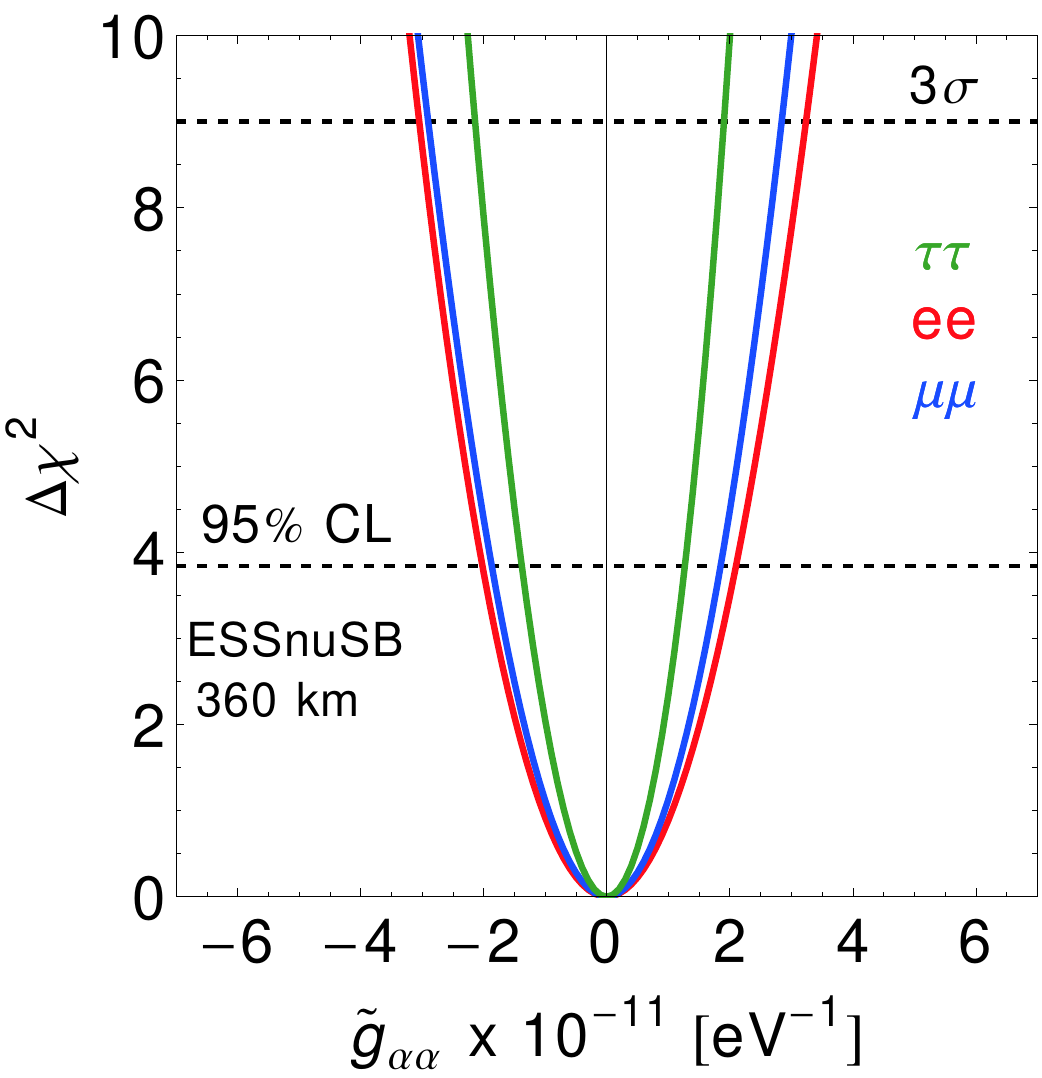}
		\end{subfigure}
		\hfill	
		 \caption{The expected sensitivity regions: SME coefficients $a_{\alpha \alpha}$ (left panel), and effective couplings $\tilde{g}_{\alpha \alpha}$ (right panel), accordingly. We marginalize over $\theta_{23}$ and $\delta_{CP}$ around their 1$\sigma$ uncertainty~\cite{deSalas:2020pgw}. All the remaining oscillation parameters are fixed to their NO best-fit values. See text for a detailed explanation.}
  \label{f2ess}
\end{figure}

In Fig.~\ref{f2ess}, we show the projected 95\% C.L. and $3\sigma$ sensitivities to the flavor conserving SME coefficients $a_{\alpha \alpha}$ and effective couplings $\tilde{g}_{\alpha \alpha}$, considering the ESSnuSB 360 km setup. The left panel displays the sensitivities to the isotropic (diagonal) SME coefficients: $a_{\alpha \alpha} \in (-4.4,~4.4) \times 10^{-23}$ GeV (red, blue, green)-solid lines ($\alpha \alpha = ee,~\mu\mu,~\tau\tau$) at 95\% C.L., we marginalized over the standard oscillation parameters $\theta_{23}$ and $\delta_{CP}$, accordingly. Moreover, the right panel shows the sensitivities to the effective couplings: $\tilde{g}_{\alpha \alpha} \in (-2.2,~2.2) \times 10^{-11}$ eV$^{-1}$ at 95\% C.L. As in the case of the non-diagonal effective couplings $\tilde{g}_{\alpha \beta}$, the expected sensitivities from the flavor-conserving sector at ESSnuSB are comparable with limits ($\tilde{g}_{\alpha \alpha} < 2 \times 10^{-11}$ eV$^{-1}$) from JUNO~\cite{Huang:2018cwo}. Besides, from Figs.~\ref{f1ess} and \ref{f2ess}, we observe that the ESSnuSB 360 km setup will be slightly more sensible to both the SME coefficients and effective couplings from the flavor changing ($e-\tau$) and flavor conserving ($\tau-\tau$) sectors.

\subsection{Sensitivity impact by improved oscillation parameter measurements}
\label{imparam}
In this subsection, we will discuss how the sensitivity to the LIV parameters $(a_L)^{T}$ may be affected by better measurements of the neutrino oscillation parameters. For example, the sensitivity may be slightly enhanced for the case of the non-diagonal $e-\mu$ sector, depending on the values of the leptonic $CP$ phase. Furthermore, if we consider the scenario of maximal mixing for the atmospheric mixing angle, which is~$\theta_{23}=45^{\circ}$, the estimated sensitivities for the $e-\mu$ and $e-\tau$ sectors may be improved correspondingly.

\begin{figure}[H]
		\begin{subfigure}[h]{0.48\textwidth}
			\caption{  }
\label{gX1}
\includegraphics[width=\textwidth]{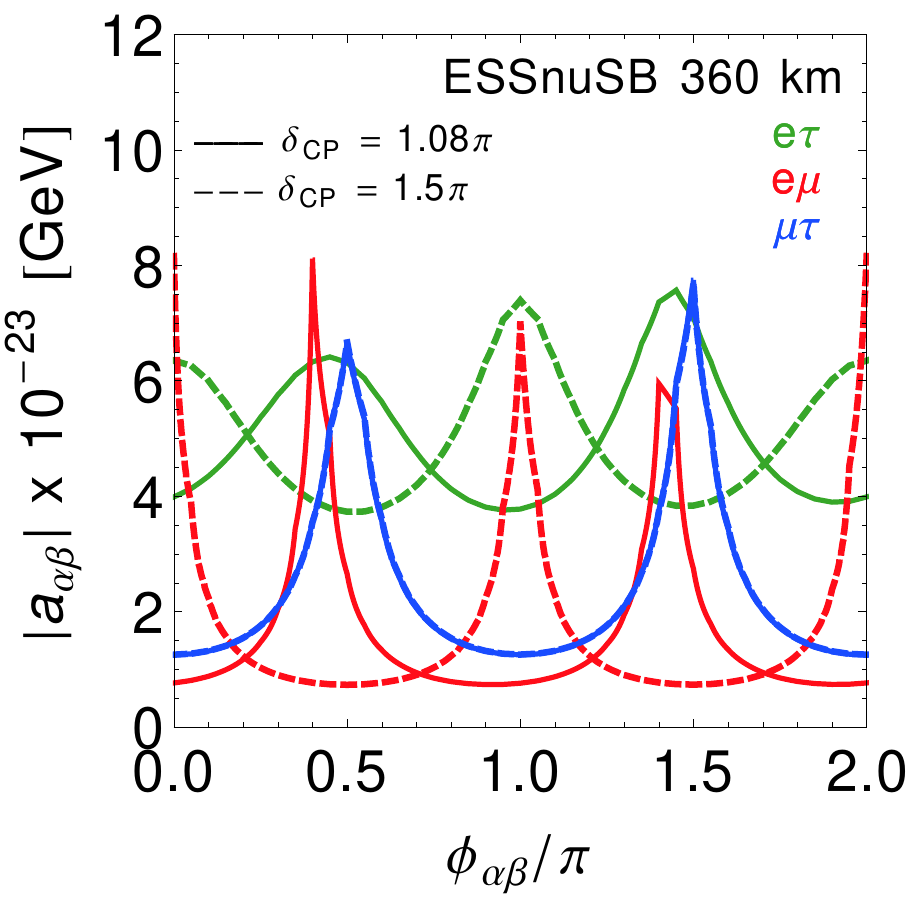}
		\end{subfigure}
		\hfill
		\begin{subfigure}[h]{0.48 \textwidth}
			\caption{}
			\label{gX2}
			\includegraphics[width=\textwidth]{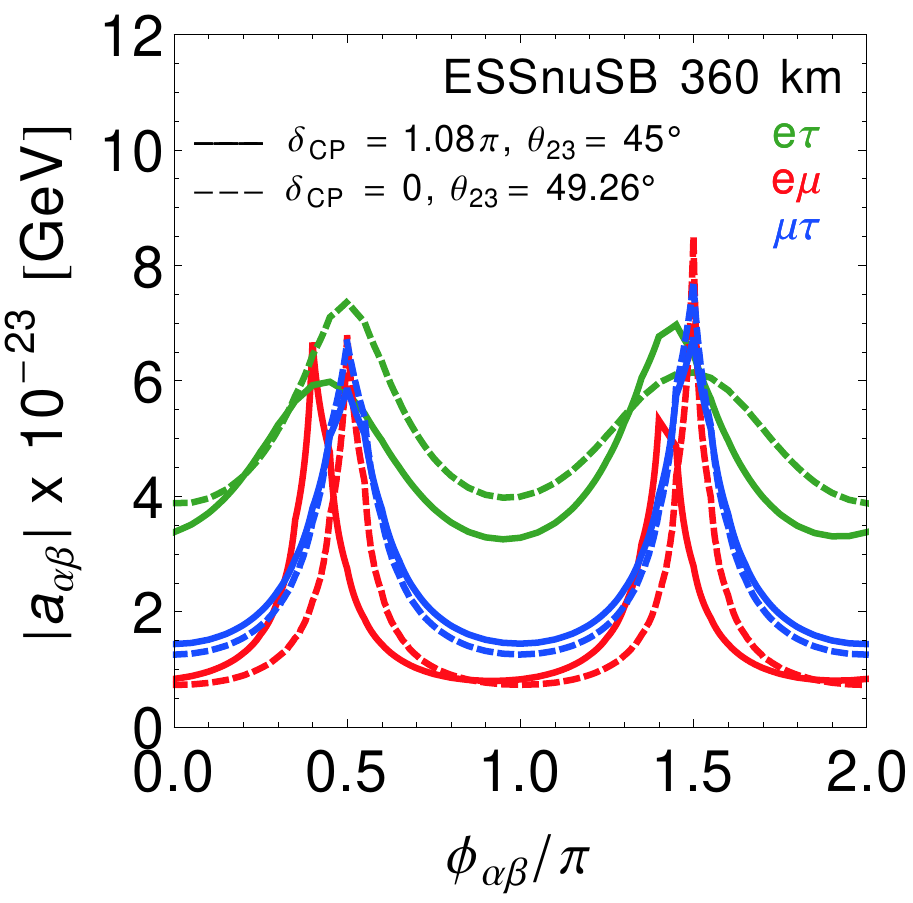}
		\end{subfigure}
		\hfill	
 \caption{Expected sensitivity regions in the $a_{\alpha \beta}$ vs.~$|\phi_{\alpha \beta}|$ projection plane. The (red, green, blue)-lines correspond to the LIV coefficients from the $\alpha \beta = (e \mu, e \tau, \mu \tau)$ sectors, respectively. The left panel considers the sensitivity influence of assuming different values of the leptonic $CP$ phase $\delta_{CP}:~(\delta_{CP}=1.08 \pi,~\delta_{CP}=1.5\pi)$ displayed as (solid, dashed)-lines. While the right panel considers the sensitivity impact in the case of two distinct values for $\theta_{23}$ and $\delta_{CP}:(\theta_{23}=45^{\circ},~\theta_{23}=49.26^{\circ}),~(\delta_{CP}=1.08 \pi,~\delta_{CP}=0)$ shown as (solid, dashed)-lines. All the remaining oscillation parameters were fixed to their NO best-fit value from Ref.~\cite{deSalas:2020pgw}.}
  \label{A1ess}
\end{figure}

In Fig.~\ref{A1ess}, we present the projected 95\% C.L. sensitivities to the isotropic non-diagonal LIV coefficients $a_{\alpha \beta} = \abs{a_{\alpha \beta}} e^{i \alpha \beta}$, considering the ESSnuSB 360 km setup. The left panel displays the projected sensitivities assuming two different values of the leptonic $CP$ phase $\delta_{CP}:~(\delta_{CP}=1.08 \pi,~\delta_{CP}=1.5\pi)$ presented as (solid, dashed)-lines. Moreover, the right panel shows the expected sensitivities in the case of two distinct values for $\theta_{23}$ and $\delta_{CP}:(\theta_{23}=45^{\circ},~\theta_{23}=49.26^{\circ}),~(\delta_{CP}=1.08 \pi,~\delta_{CP}=0)$ displayed as (solid, dashed)-lines. All the remaining oscillation parameters are fixed to their NO best-fit value, shown in Table~\ref{tab:1}. We observe that the projected sensitivities to the non-diagonal LIV parameters will be almost unaffected by assuming distinct values of $\delta_{CP}$, exhibiting a mild sensitivity enhancement from the $e-\mu$ sector in the case of $\delta_{CP}=0$, as displayed in the right panel of Fig.~\ref{A1ess}. While the sensitivities to the other non-diagonal sectors are not modified in comparison with our benchmark value of $\delta_{CP} = 1.08 \pi$. As a result, the effect of considering distinct values of the $CP$ phase does not have a considerable influence for this particular type of new physics. A similar discussion follows for the case of the effective neutrino$-$UALP couplings $\tilde{g}_{\alpha \beta}$.

\begin{figure}[H]
		\begin{subfigure}[h]{0.48\textwidth}
			\caption{  }
\label{FX1}
\includegraphics[width=\textwidth]{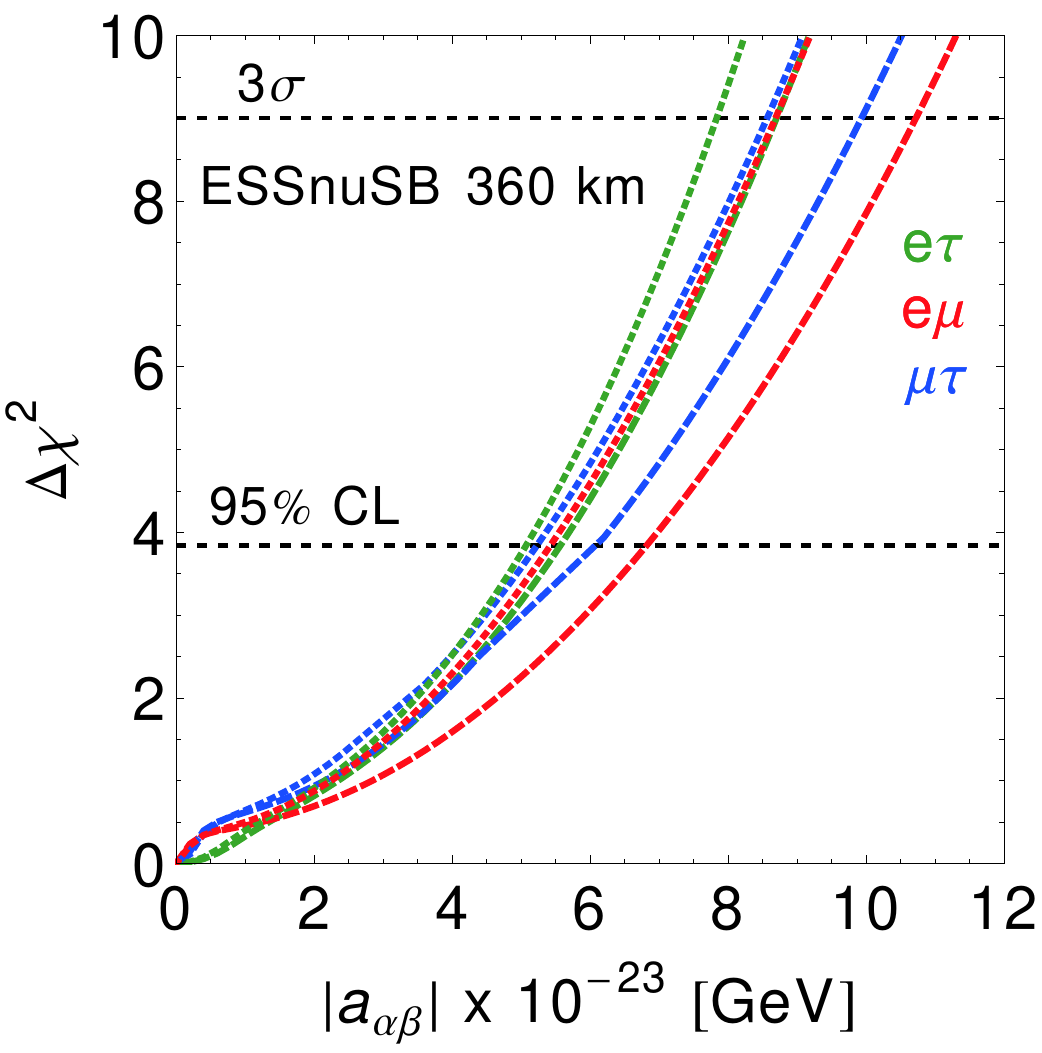}
		\end{subfigure}
		\hfill
		\begin{subfigure}[h]{0.48 \textwidth}
			\caption{}
			\label{FX2}
			\includegraphics[width=\textwidth]{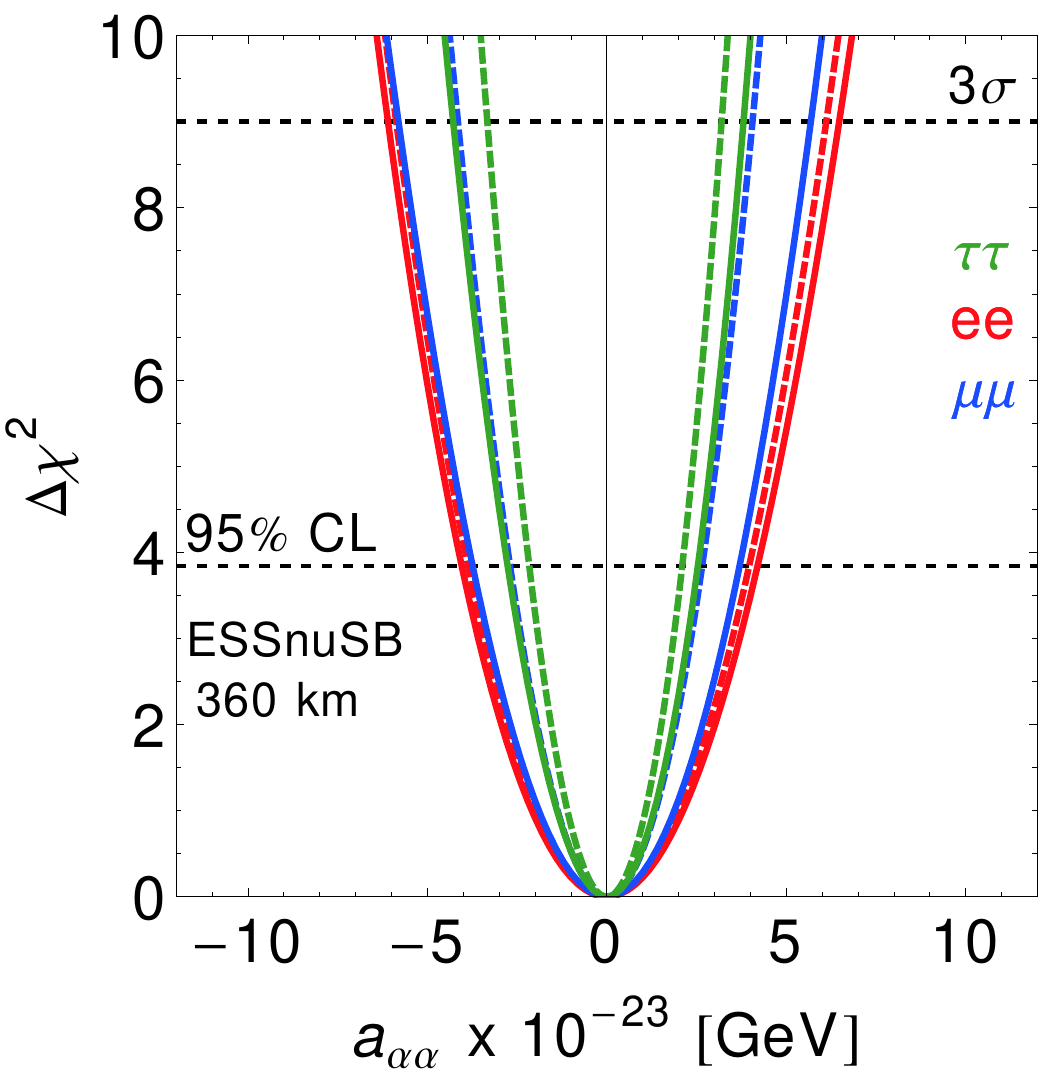}
		\end{subfigure}
		\hfill	
 \caption{Expected sensitivity regions in the $|a_{\alpha \beta}|$ (left panel) and $a_{\alpha \alpha}$ (right panel) projection planes, accordingly. The (red, green, blue)-lines correspond to the LIV coefficients from the $\alpha \beta = (e \mu, e \tau, \mu \tau)$ and $\alpha \alpha = (e e, \tau \tau, \mu \mu)$ sectors, respectively. We marginalize over $\theta_{23}$ and ($\delta_{CP}$) assuming future 1$\sigma$ uncertainties of 3\% and (5\%), accordingly. For the non-diagonal LIV coefficients, we marginalized over the corresponding LIV phases, $\phi_{\alpha \beta}$, in the full [0$-2\pi$] range. All the remaining oscillation parameters were fixed to their NO best-fit value from~\cite{deSalas:2020pgw}. Refer to the text for a detailed explanation.}
  \label{A2ess}
\end{figure}

In Fig.~\ref{A2ess}, we consider the case of future improvements on the measurements of the neutrino mixing parameters $\theta_{23}$ and $\delta_{CP}$, accordingly.~\footnote{At present, out of the complete set of neutrino oscillation mixing parameters, the oscillation parameters $\theta_{23}$ and $\delta_{CP}$ are the largest sources of uncertainties~\cite{deSalas:2020pgw}. The current uncertainty on the measurement of the reactor mixing angle $\theta_{13}$ is at the percent level, while for the solar mixing parameters ($\Delta m_{21}^2, \theta_{12}$), particularly the solar mixing angle $\theta_{12}$ uncertainty is considerably. These mixing parameters play a sub-leading role in accelerator-based oscillation experiments, such as the ESSnuSB proposal. As far as the atmospheric mass-squared difference $\Delta m_{31}^2$ is concerned, we fix it to its present NO best-fit value~\cite{deSalas:2020pgw}.} In order to examine such impacts, we will assume a future 1$\sigma$ uncertainty of 3\% for the atmospheric mixing angle $\theta_{23}$ and 5\% for the case of the leptonic $CP-$violating phase $\delta_{CP}$. The left panel displays the projected sensitivities to the non-diagonal LIV coefficients $|a_{\alpha \beta}|$; dotted lines correspond to $\theta_{23} = 45^{\circ}$, while dashed lines consider our benchmark values from Table~\ref{tab:1}. In contrast, the right panel shows the expected sensitivities to the diagonal LIV coefficients $a_{\alpha \alpha}$; dashed lines consider $\theta_{23} = 45^{\circ}$, while solid lines assume our standard oscillation values from Table~\ref{tab:1}.

From the left panel of Fig.~\ref{f1ess} and Fig.~\ref{A2ess} (dashed lines), we observe that for our standard benchmark values, the impact of improved oscillation parameter measurements on the future sensitivities to the non-diagonal LIV coefficients $|a_{\alpha \beta}|$ is marginal. However, if we consider the case with $\theta_{23} = 45^{\circ}$ (dotted lines), there could be a considerable enhancement of the sensitivities from the $e-\mu$ and $\mu-\tau$ sectors, while the sensitivity in the $e-\tau$ sector shows only a mild improvement.
Besides, in the case of the diagonal LIV coefficients $a_{\alpha \alpha}$ (right panel of Fig.~\ref{A2ess}), only a significant sensitivity improvement may be expected from the $\mu-\mu$ sector. A similar discussion applies to the case of the effective neutrino$-$UALP couplings: $\tilde{g}_{\alpha \beta}$ and $\tilde{g}_{\alpha \alpha}$.

Regarding astrophysical limits on the effective neutrino$-$UALP couplings $\tilde{g}_{\alpha \beta}$, the most restrictive constraints originate from the decay process: $\nu_{\alpha} \leftrightarrow \nu_{\beta} + \varphi$, which establishes $\tilde{g}_{\alpha \beta} \lesssim 10^{-10}$ eV$^{-1}$~\cite{Huang:2018cwo}. For a summary of current limits and projected sensitivities to the isotropic SME coefficients ($a_{\alpha \beta}$, $a_{\alpha \alpha}$) and the effective neutrino$-$UALP couplings ($\tilde{g}_{\alpha \beta}$, $\tilde{g}_{\alpha \alpha}$) at ESSnuSB 360 km, and other experimental configurations, see Table~\ref{tab:2}.

\begin{table}[H] 
\caption{\label{tab:2} Summary of current limits and projected sensitivities (shown in parenthesis) from beyond the standard model: neutrino LIV, neutrino$-$UALP interaction, and neutrino$-$UALP decay processes relevant to this study.}
\centering
\begin{tabular}{ c    c }
\hline \hline
~BSM process~ & Limit (Future sensitivity) \\
\hline 
~Neutrino LIV ~~&~~~$|a_{e \mu}| < 1.8 \times 10^{-23}$ GeV,~Super-Kamiokande~\cite{Super-Kamiokande:2014exs}~~~\\ 
~Neutrino LIV ~~&~~~$|a_{e \tau}| < 2.8 \times 10^{-23}$ GeV,~Super-Kamiokande~\cite{Super-Kamiokande:2014exs}~~~\\
~Neutrino LIV ~~&$|a_{\mu \tau}| < 2.9 \times 10^{-24}$ GeV,~IceCube~\cite{IceCube:2017qyp} \\
~Neutrino LIV ~~&$a_{\tau \tau} < 2 \times 10^{-26}$ GeV,~IceCube~\cite{Arguelles:2024cjj} \\
~Neutrino LIV ~~&($|a_{\alpha \beta}| \simeq 1 \times 10^{-23}$ GeV),~DUNE~\cite{Barenboim:2018ctx, Agarwalla:2023wft, Raikwal:2023lzk}  \\
~Neutrino LIV~~&($|a_{\alpha \beta}| \simeq 5 \times 10^{-23}$ GeV),~Hyper-K~\cite{Agarwalla:2023wft} \\
~Neutrino LIV~~&$(|a_{\alpha \beta}| \simeq [1-4] \times 10^{-23}$ GeV),~ICAL~\cite{Raikwal:2023lzk}\\
~Neutrino LIV~~&$(|a_{\alpha \beta}| \simeq [2-7] \times 10^{-23}$ GeV),~T2HK~\cite{Raikwal:2023lzk}\\
~Neutrino LIV~~&($a_{\alpha \beta} \lesssim 10^{-30}$ GeV),~UHE neutrinos~\cite{Ando:2009ts, Klop:2017dim}\\
~Neutrino LIV ~~&$(|a_{\alpha \beta}| \simeq [5-7] \times 10^{-23}$ GeV),~ESSnuSB 360 km\\
~Neutrino LIV ~~&~$(a_{\alpha \alpha} \in (-4.4,~4.4) \times 10^{-23}$ GeV),~ESSnuSB 360 km\\
~~Neutrino-UALP  ~~&~($\tilde{g}_{\alpha \beta} \simeq 10^{-11}-10^{-12}$ eV$^{-1}$),~DUNE~\cite{Huang:2018cwo} \\
~~Neutrino-UALP  ~~&~$\tilde{g}_{\tau \tau} < 3 \times 10^{-13}$ eV$^{-1}$,~IceCube~\cite{Arguelles:2024cjj} \\
~~Neutrino-UALP  ~~&~$\tilde{g}_{\alpha \beta} \lesssim 10^{-13}$ eV$^{-1}$~\cite{Cordero:2023hua} \\
~~Neutrino-UALP  ~~&~$(\tilde{g}_{\alpha \beta} < 2 \times 10^{-11}$ eV$^{-1})$,~JUNO~\cite{Huang:2018cwo} \\
~~Neutrino-UALP  ~~&~$(|\tilde{g}_{\alpha \beta}| \simeq 3\times 10^{-11}$ eV$^{-1}$),~ESSnuSB 360 km \\
~~Neutrino-UALP  ~~&~$(\tilde{g}_{\alpha \alpha} \in (-2.2,~2.2) \times 10^{-11}$ eV$^{-1}$),~ESSnuSB 360 km~~\\
$\nu_{\alpha} \leftrightarrow \nu_{\beta} + \varphi$~&~$\tilde{g}_{\alpha \beta} \lesssim 10^{-10}$ eV$^{-1}$~\cite{Huang:2018cwo}\\
 \hline \hline
\end{tabular}
\end{table}


\section{Conclusion}
\label{conclusion}

The dark matter puzzle stands out as one of the main challenges to understand in both particle physics and cosmology. Ultralight axion-like dark matter (UALP), is an appealing dark matter candidate that can be searched for in neutrino oscillation experiments. Besides, this type of neutrino$-$UALP interaction, could induce breakdowns of the Lorentz and $CPT$ symmetries in the neutrino sector. At the cosmological level, the ALP can act as DM, this is because the neutrino$-$UALP energy exchange term is $Q \simeq 0$, as described in Section~\ref{cosmo}. 

In this paper, we have studied the sensitivity to the isotropic $CPT-$odd SME coefficients $(a_L)^{T}$, and the effective couplings ($\tilde{g}$) of the neutrino-ultralight ALP interaction, at the next-to-next generation long-baseline neutrino experiment ESSnuSB, our results are summarized in Table~\ref{tab:2}. However, improved measurements of the neutrino oscillation parameters, such as the atmospheric mixing angle $\theta_{23}$, may enhance the projected sensitivity of the LIV parameters $(a_L)^{T}$. Therefore, the ESSnuSB configuration represents an appealing platform to investigate both LIV and potential neutrino$-$UALP interactions. For instance, the expected sensitivity at the ESSnuSB 360 km configuration is comparable with limits from JUNO~\cite{Huang:2018cwo}. 
It would be interesting to study the complementary among the upcoming ESSnuSB and JUNO experiments within this scenario.


\section*{Acknowledgments}
We would like to thank Omar Miranda for useful discussions. We acknowledge the anonymous referee for the illuminating comments and suggestions. This work was partially supported by SNII-M\'exico and CONAHCyT research Grant No.~A1-S-23238. Additionally, the work of R.~C. was partially supported by COFAA-IPN, Estímulos al Desempeño de los Investigadores (EDI)-IPN and SIP-IPN Grant No.~20241624. 


\end{document}